\documentclass[preprint,amsmath,amssymb,aps,nofootinbib,showpacs]{revtex4}
\usepackage{graphicx}
\usepackage{bm}
\usepackage{subfigure}
\usepackage{amsmath}

\newcommand{\pr}{Phys. Rev.\ }
\newcommand{\pla}{Phys. Lett. A\ }
\newcommand{\jpa}{J. Phys. A\ }
\newcommand{\jpb}{J. Phys. B\ }
\newcommand{\etal}{{\em et al. }}

\newcommand{\UQ}{School of Mathematics and Physics, University of Queensland, Brisbane, 
QLD 4072, Australia.}
\newcommand{\Otago}{Jack Dodd Centre for Quantum Technology, Department of Physics, University of Otago, Dunedin, New Zealand.}

\begin{document}

\title{Quantum ultra-cold atomtronics}

\author{M.~K. Olsen}
\affiliation{\UQ}
\author{A.~S. Bradley}
\affiliation{\Otago}
%\author{C.~V. Chianca}
%\affiliation{\UQ}
%-----------------------------------------------------------------------
\date{\today}
%------------------------------------------------------------------------

\begin{abstract}

It is known that a semi-classical analysis is not always adequate for atomtronics devices, but that a fully quantum analysis is often necessary to make reliable predictions. While small numbers of atoms at a small number of sites are tractable using the density matrix, a fully quantum analysis is often not straightforward as the system becomes larger. We show that the fully quantum positive-P representation is then a viable calculational tool. We postulate an atomtronic phase-gate consisting of four wells in a Bose-Hubbard configuration, for which the semi-classical dynamics are controllable using the phase of the atomic mode in one of the wells. We show that the quantum predictions of the positive-P representation for the performance of this device have little relation to those found semi-classically, and that the performance depends markedly on the actual quantum states of the initially occupied modes. We find that initial coherent states lead to closest to classical dynamics, but that initial Fock states give results that are quite different. A fully quantum analysis also opens the door for deeply quantum atomtronics, in which properties such as entanglement and EPR (Einstein-Podolsky-Rosen) steering become valuable technical properties of a device.

\end{abstract}
%******************************************* 

       % check these

\maketitle

\section{Introduction}
\label{sec:intro}

Atomtronics is an emerging area of investigation in which analogues of electronic circuits and devices are constructed using ultra-cold bosonic atoms rather than electrons as in conventional electronics~\cite{overview}. The conventional way to construct an atomtronic device is to use cold atoms trapped in an optical lattice, which has a description in terms of either the Mott-Hubbard model transferred from condensed matter physics~\cite{MHmodel1,MHmodel2,MHmodel3} for fermionic atoms, and the Bose-Hubbard model for bosonic atoms~\cite{BHmodel}. These models can represent either insulating behaviour, in the Mott insulator regime, or conducting behaviour, in the superfluid regime. In this work we consider only bosonic atoms. Shortly after the realisation of trapped Bose-Einstein condensates (BEC), Jaksch \etal~\cite{Jaksch} showed that the Bose-Hubbard model can provide an accurate description of bosonic atoms trapped in a deep optical lattice. 
The basics of this model have been used to investigate a wide variety of atomtronic devices~\cite{overview}, including one with diode-like behaviour~\cite{Ruschhaupt}, a single-atom switching transistor~\cite{Micheli}, and one which uses a triple well configuration to mimic a field effect transistor~\cite{Stickney}. More recent proposals include circuits of diodes and transistors~\cite{Pepino}, and an atomtronic battery~\cite{battery}. Gajdacz \etal proposed atomtronic transistors with the idea of combining them into gates for quantum computers~\cite{Gajdacz}. 
We have no doubt that there will be more proposals in the future, and what we will show here is that a full quantum description of both the dynamics and the initial state needs to be taken into account to guarantee accurate descriptions of the dynamics of such devices.

Various theoretical methods have been used to analyse these devices up until the present. For small numbers of atoms and sites, direct quantum calculations using a master equation are often possible. 
Pepino \etal  developed a quantum master equation to treat systems which interact with sources and sinks, based largely on methods which have been extremely successful in quantum optics. With this method, they have analysed electronic diodes, field-effect transistors, bipolar junction transistors, and an analogy to a logic gate~\cite{open}. The master equation was then solved numerically, giving fully quantum solutions. This method allows for the investigation of systems with relatively small numbers of atoms and lattice sites, with the authors typically treating $3$ or $4$ sites, each with $1$ or $2$ atoms. Gajdacz \etal used coupled nonlinear  Schr\"{o}dinger equations in one dimension to calculate the phase evolution of eigenstates, with two distinguishable atoms in a triple well~\cite{Gajdacz}. For the small systems considered by these authors, their methods are perfectly adequate.
In the mesoscopic regime, when the numbers of atoms can make density matrix calculations complicated, the one-body Schr\"{o}dinger equation can be used for non-interacting atoms, or the mean-field Gross-Pitaevskii equations (GPE) for interacting atoms. Both of these have been used to model coherent atomic transport in a three-well potential~\cite{Rab2008,Bradly2012,Opatrny2009} and each has disadvantages. The one-body Schr\"{o}dinger equation, being linear, cannot include the interactions required for the Mott insulator regime. The GPE approach cannot describe any quantum statistical features, such as the actual quantum states or any entanglement properties. The three-well coherent transport model has previously been analysed using the fully quantum positive-P representation~\cite{Pplus}, which was used to note the differences caused by different initial quantum states on the dynamics, the entanglement properties~\cite{myJPB} and the quantum steering properties~\cite{myJOSAB}.

In what follows we base ourselves on the approach taken by Milburn \etal~\cite{BHJoel}, generalisng this to four wells~\cite{Chianca4well,Chiancathermal}, and using the fully quantum positive-P phase space representation. We consider this to be the most suitable approach here because it is exact, allows for an easy representation of mesoscopic numbers of atoms, and can simulate different quantum states~\cite{states}. Just as importantly, the positive-P calculations scale linearly with the number of sites and can in principle deal with any number of atoms.
These are powerful advantages when we wish to consider mesoscopic numbers of atoms in arbitrary numbers of potential wells. One disadvantage of the positive-P representation is that the integration has a tendency to diverge for high collisional nonlinearities~\cite{Steel}, although it often converges for long enough to show marked differences from 
mean-field predictions~\cite{JoeSuper} and also allows for the calculation of quantum correlations~\cite{EPRBEC}.

\section{Physical model, Hamiltonian and equations of motion}
\label{sec:model}

\begin{figure}
\begin{center}
\includegraphics[width=0.8\columnwidth]{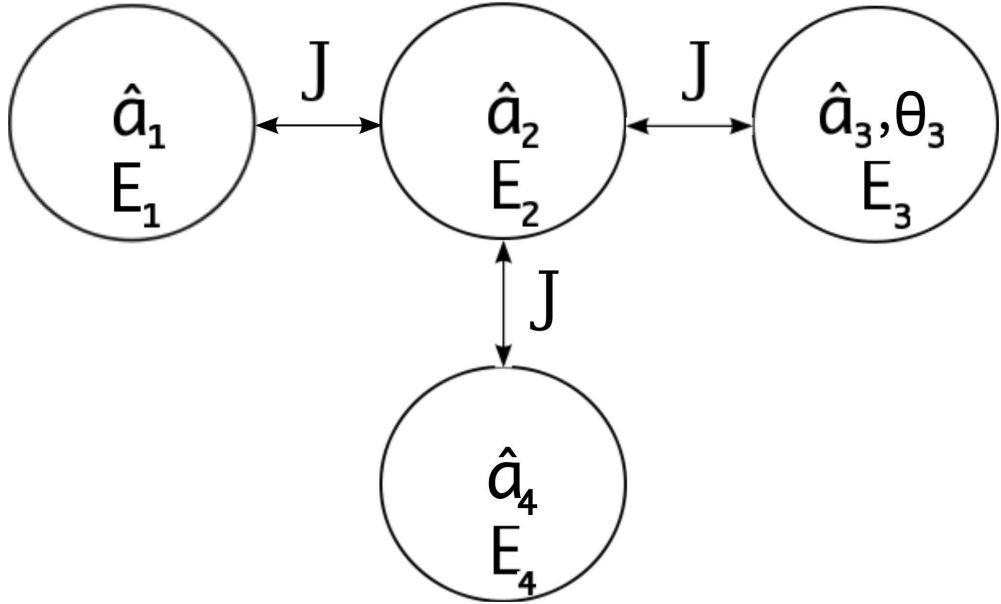}
\end{center}
\caption{(Color online) A simplified schematic of the device. Wells $1,2$ and $3$ have nearest neighbour couplings, and well $4$ is coupled with well $2$. The $\hat{a}_{j}$ are bosonic annihilation operators for atoms in mode $j$, the $E_{j}$ are the ground state single-atom energies of the wells, and $J$ represents the coupling between the wells. $\theta_{3}$ represents a phase-shift which is applied to mode $3$ at the beginning of the evolution, when the populations of $1$ and $3$ will be equal, with $2$ and $4$ initially being empty.}
\label{fig:setup}
\end{figure}

The four-well system is as shown in the schematic of Fig.~\ref{fig:setup}, where the circles represent the wells or lattice sites, each of which contains a single atomic mode. The $\hat{a}_{j}$ are bosonic annihilation operators for atoms in mode $j$, the $E_{j}$ are the ground state single-atom energies of the wells, and $J$ represents the coupling between the wells. We assume that there is no coupling between well $4$ and wells $1$ and $3$. We also assume that any atoms at each site are initially in their ground state. The basic idea is that atoms from the outside wells will tunnel into the centre well and from there to well $4$.  Because of the wave nature of the condensed atoms, we expect that the initial phase difference, $\theta_{3}$ between the atomic modes at sites $1$ and $3$ will affect the rate of tunnelling into the middle and hence into well $4$. We would hence have a type of phase sensitive gate, where the phase of one mode mode can be used to control the occupation of another. 

Following the usual procedures~\cite{BHJoel}, with $\chi$ as the s-wave collisional term, we write the Hamiltonian as
\begin{eqnarray}
{\cal H} = & & \hbar\sum_{j}E_{j}\hat{a}_{j}^{\dag}\hat{a}_{j}+\hbar\sum_{j}\chi \hat{a}_{j}^{\dag\;2}\hat{a}_{j}^{2}\nonumber\\
& & -\hbar J\left(\hat{a}_{1}^{\dag}\hat{a}_{2}+\hat{a}_{2}^{\dag}\hat{a}_{1}+\hat{a}_{3}^{\dag}\hat{a}_{2}
+\hat{a}_{2}^{\dag}\hat{a}_{3}
+\hat{a}_{4}^{\dag}\hat{a}_{2}+\hat{a}_{2}^{\dag}\hat{a}_{4}\right),
\label{eq:Ham}
\end{eqnarray}
where $j$ runs from $1$ to $4$. Starting from this Hamiltonian, our first step is to find the semi-classical mean-field equations in the Gross-Pitaevskii approach. We will use the solutions of these for comparison purposes, since it is well known that they are not always accurate for the Bose-Hubbard model, even for calculation of the 
mean fields~\cite{BHJoel,Chianca4well,Chiancathermal}. Using the variables $\alpha_{j}$ to represent the mean-fields, we find
\begin{eqnarray}
\frac{d\alpha_{1}}{dt} &=& -i\left(E_{1}+2\chi |\alpha_{1}|^{2}\right)\alpha_{1}+iJ\alpha_{2},\nonumber\\
\frac{d\alpha_{2}}{dt} &=& -i\left(E_{2}+2\chi |\alpha_{2}|^{2}\right)\alpha_{2}+iJ\left(\alpha_{1}+\alpha_{3}+\alpha_{4}\right),\nonumber\\
\frac{d\alpha_{3}}{dt} &=& -i\left(E_{3}+2\chi |\alpha_{3}|^{2}\right)\alpha_{3}+iJ\alpha_{2},\nonumber\\
\frac{d\alpha_{4}}{dt} &=& -i\left(E_{4}+2\chi |\alpha_{4}|^{2}\right)\alpha_{4}+iJ\alpha_{2}.
\label{eq:GPE}
\end{eqnarray}

\begin{figure}
\begin{center}
\includegraphics[width=0.8\columnwidth]{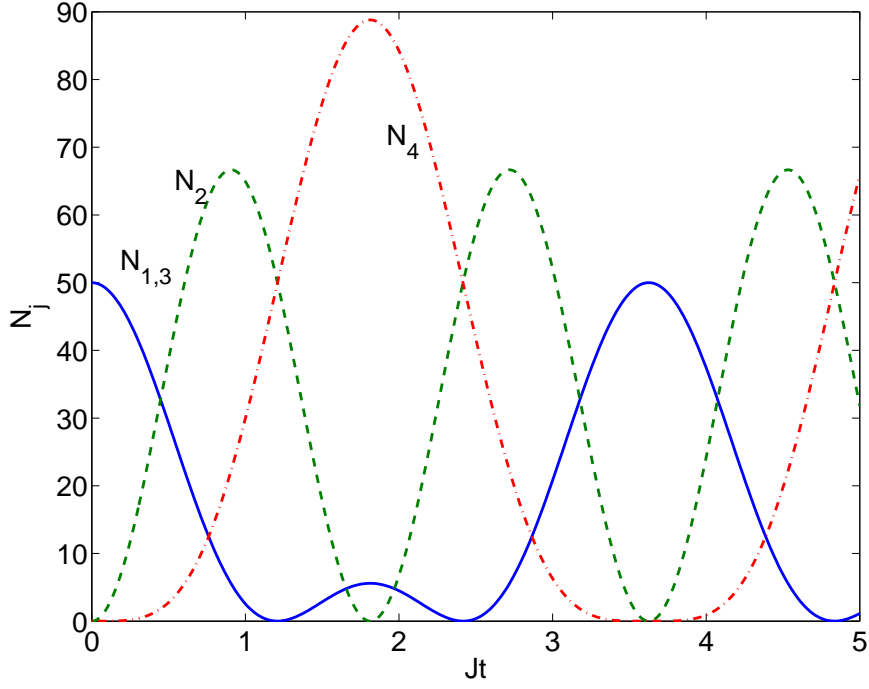}
\end{center}
\caption{(Color online) Solutions of Eq.~\ref{eq:GPE} for the numbers in each well as a function of dimensionless time, $Jt$. The initial phase difference, $\theta_{3}$, between wells $1$ and $3$ is zero, and the numbers in these two wells are equal. Wells $2$ and $4$ are initially empty. We see that atoms pass through the middle well and populate the fourth. The parameters used in this and all subsequent plots are $E_{j}=0 \forall j$, $N_{1}(0)=N_{3}(0)=50$, $N_{2}(0)=N_{4}(0)=0$, $J=1$, and $\chi=10^{-3}$.}
\label{fig:classzero}
\end{figure}

To solve the full quantum equations, we use the positive-P representation~\cite{Pplus}, which allows for exact solutions of the dynamics arising from the Hamiltonian of Eq.~\ref{eq:Ham} in terms of normally-ordered operator averages, in the limit of the average of an infinite number of trajectories of stochastic differential equations in a doubled phase-space. This method also allows for the representation of different quantum states in the initial conditions~\cite{states}. The positive-P representation does have the disadvantage that it can suffer from divergence problems in cases with a largish colliisional nonlinearity, often limiting its utility to short time dynamics~\cite{Steel}. In this letter, it is exactly this regime we are analysing, so that the positive-P representation is perfectly adequate. Following the standard methods~\cite{DFW}, the set of It\^o stochastic differential equations~\cite{SMCrispin} are found as
\begin{eqnarray}
\frac{d\alpha_{1}}{dt} &=& -i\left(E_{1}+2\chi\alpha_{1}^{+}\alpha_{1}\right)\alpha_{1}+iJ\alpha_{2}
+\sqrt{-2i\chi\alpha_{1}^{2}}\;\eta_{1},\nonumber\\
\frac{d\alpha_{1}^{+}}{dt} &=& i\left(E_{1}+2\chi\alpha_{1}^{+}\alpha_{1}\right)\alpha_{1}^{+}-iJ\alpha_{2}^{+}
+\sqrt{2i\chi\alpha_{1}^{+\;2}}\;\eta_{2},\nonumber\\
\frac{d\alpha_{2}}{dt} &=& -i\left(E_{2}+2\chi\alpha_{2}^{+}\alpha_{2}\right)\alpha_{2}+iJ\left(\alpha_{1}
+\alpha_{3}+\alpha_{4}\right)
+\sqrt{-2i\chi\alpha_{2}^{2}}\;\eta_{3},\nonumber\\
\frac{d\alpha_{2}^{+}}{dt} &=& i\left(E_{2}+2\chi\alpha_{2}^{+}\alpha_{2}\right)
\alpha_{2}^{+} -iJ\left(\alpha_{1}^{+}
+\alpha_{3}^{+}+\alpha_{4}^{+}\right)
+\sqrt{2i\chi\alpha_{2}^{+\,2}}\;\eta_{4},\nonumber\\
\frac{d\alpha_{3}}{dt} &=& -i\left(E_{3}+2\chi\alpha_{3}^{+}\alpha_{3}\right)\alpha_{3}+iJ\alpha_{2}
+\sqrt{-2i\chi\alpha_{3}^{2}}\;\eta_{5},\nonumber\\
\frac{d\alpha_{3}^{+}}{dt} &=& i\left(E_{3}+2\chi\alpha_{3}^{+}\alpha_{3}\right)\alpha_{3}^{+}-iJ\alpha_{2}^{+}
+\sqrt{2i\chi\alpha_{3}^{+\;2}}\;\eta_{6},\nonumber\\
\frac{d\alpha_{4}}{dt} &=& -i\left(E_{4}+2\chi\alpha_{4}^{+}\alpha_{4}\right)\alpha_{4}+iJ\alpha_{2}
+\sqrt{-2i\chi\alpha_{4}^{2}}\;\eta_{7},\nonumber\\
\frac{d\alpha_{4}^{+}}{dt} &=& i\left(E_{4}+2\chi\alpha_{4}^{+}\alpha_{4}\right)\alpha_{4}^{+}-iJ\alpha_{2}^{+}
+\sqrt{2i\chi\alpha_{4}^{+\;2}}\;\eta_{8},
\label{eq:Pplus}
\end{eqnarray}
where the $\eta_{j}$ are standard Gaussian noises with $\overline{\eta_{j}}=0$ and $\overline{\eta_{j}(t)\eta_{k}(t')}=\delta_{jk}\delta(t-t')$. As always, averages of the positive-P variables represent normally ordered operator moments, such that, for example, $\overline{\alpha_{j}^{m}\alpha_{k}^{+\,n}}\rightarrow\langle\hat{a}^{\dag\,n}\hat{a}^{m}\rangle$.

\section{Results}
\label{sec:results}

\begin{figure}
\begin{center}
\includegraphics[width=0.8\columnwidth]{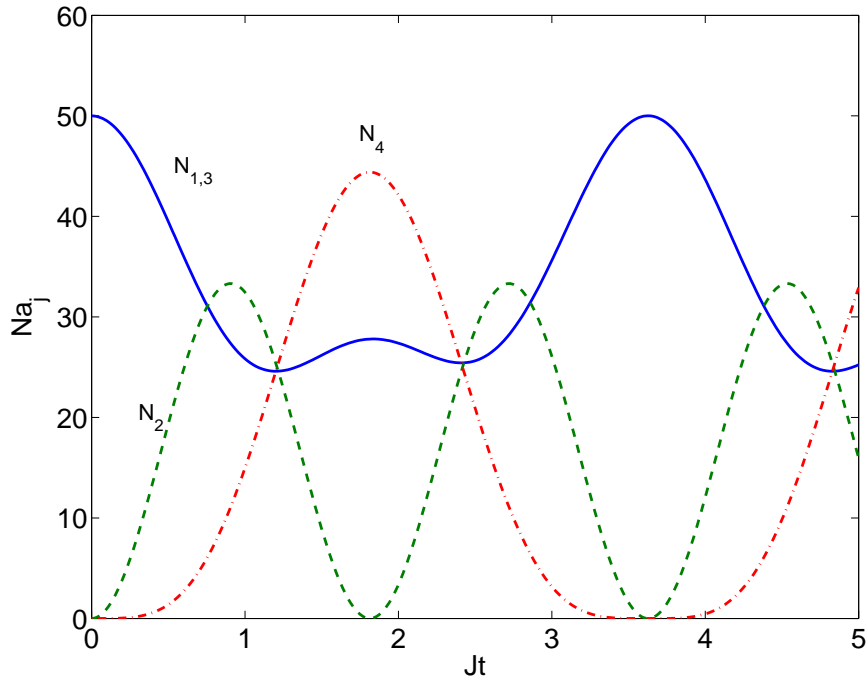}
\end{center}
\caption{(Color online) Solutions of Eq.~\ref{eq:GPE} for the numbers in each well for the same parameters and initial conditions as Fig.~\ref{fig:classzero} except that $\theta_{3}=\pi/2$. We see that the maximum population of well four is approximately half that seen in Fig.~\ref{fig:classzero}.}
\label{fig:classpihalf}
\end{figure}

In order to analyse and compare the effects of different initial states, we begin with a total of $100$ atoms, evenly distributed between wells $1$ and $3$ and analyse the classical evolution given by Eq.~\ref{eq:GPE}. The main quantity of interest is $N_{4}(t)$, the number of atoms in well $4$ as a function of time. With the initial phases equal, we see a $90\%$ transfer of atoms into the fourth well at certain times, as shown in Fig.~\ref{fig:classzero}. When we set $\theta_{3}$ to $\pi$, this fourth well remains unpopulated, showing that the device can act semi-classically as a phase-dependent gate. With $\theta_{3}=\pi/2$, as seen in Fig.~\ref{fig:classpihalf}, we see that a little less than half the atoms appear in the fourth well. Having thus benchmarked the device, we will now proceed to see how it performs with initial coherent and Fock states in wells $1$ and $3$.

The initial quantum states are modelled as in ref.~\cite{states}, with the positive-P representation equations being numerically integrated over a large number of stochastic trajectories. The immediate difference is that, because we can calculate close approximations to normally-orderd operator expectation values, we are also able to calculate the standard deviations about the average solutions for the atom numbers. We begin with coherent states with the same average occupation numbers as used classically. The important differences are that the initial physical number distribution is now Poissonian (close to Gaussian as the number of quanta increases) and quantum noise is included in the calculations. We immediately find that the average solutions for the intensities are the same as the semi-classical mean-field solutions, for both $\theta_{3}=0$ and $\theta_{3}=\pi$. The only difference is that there is a distribution about these solutions, as shown by the lines plotted at plus and minus one standard deviation. Thus the semi-classical treatment gives the most likely solutions, although we must remember that most individual experiments will give something different. How important this is in the case of initial coherent states will depend on how precisely we wish a particular atomtronic device to function.

\begin{figure}
\begin{center}
\includegraphics[width=0.8\columnwidth]{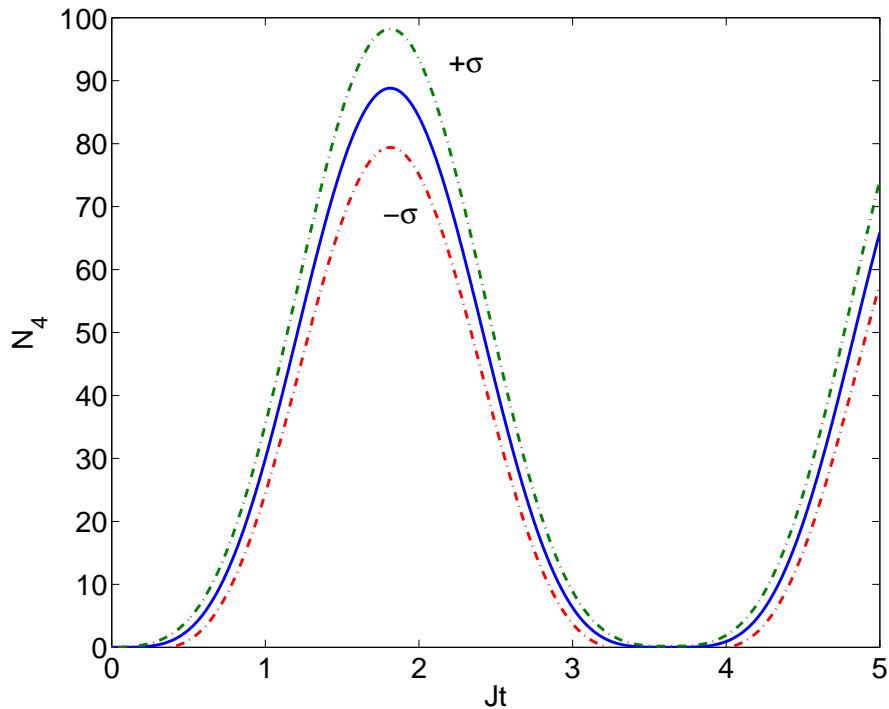}
\end{center}
\caption{(Color online) Quantum solutions for the number of atoms in the fourth well, beginning with coherent states, and the mean plus or minus one standard deviation. The initial conditions are the same as for Fig.~\ref{fig:classzero} and the solutions are the average of $6.6\times 10^{5}$ stochastic trajectories of the positive-P equations.}
\label{fig:Czero}
\end{figure}

\begin{figure}
\begin{center}
\includegraphics[width=0.8\columnwidth]{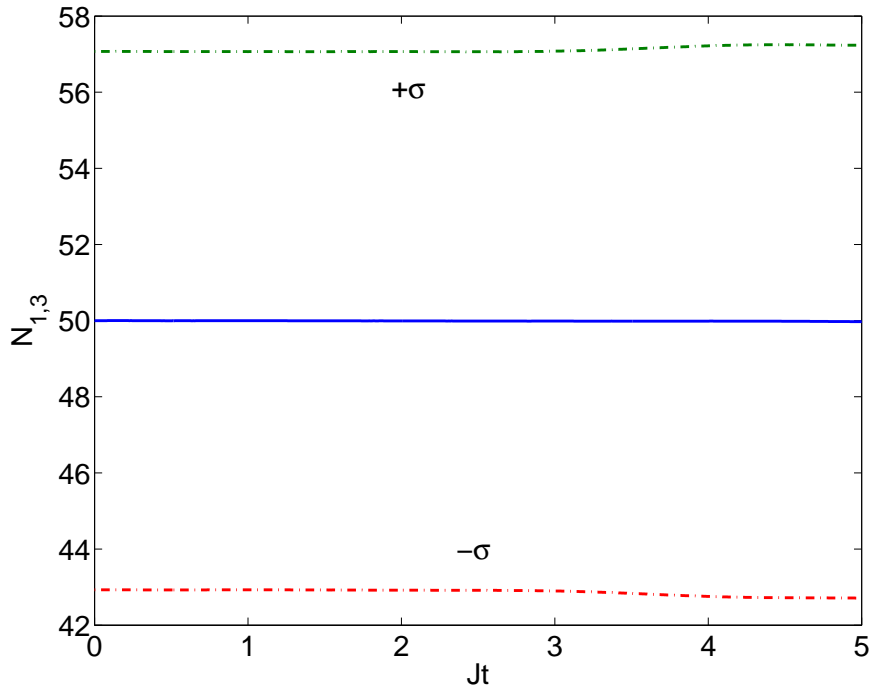}
\end{center}
\caption{(Color online) Quantum solutions for the number of atoms in the first and third wells, beginning with coherent states, and the mean plus or minus one standard deviation. $N_{1}$ and $N_{3}$ are equal. The initial conditions are the same as for Fig.~\ref{fig:Czero}, but with $\theta_{3}=\pi$, and the solutions are the average of $6.45\times 10^{5}$ stochastic trajectories of the positive-P equations.}
\label{fig:Cpi}
\end{figure}

\begin{figure}
\begin{center}
\includegraphics[width=0.8\columnwidth]{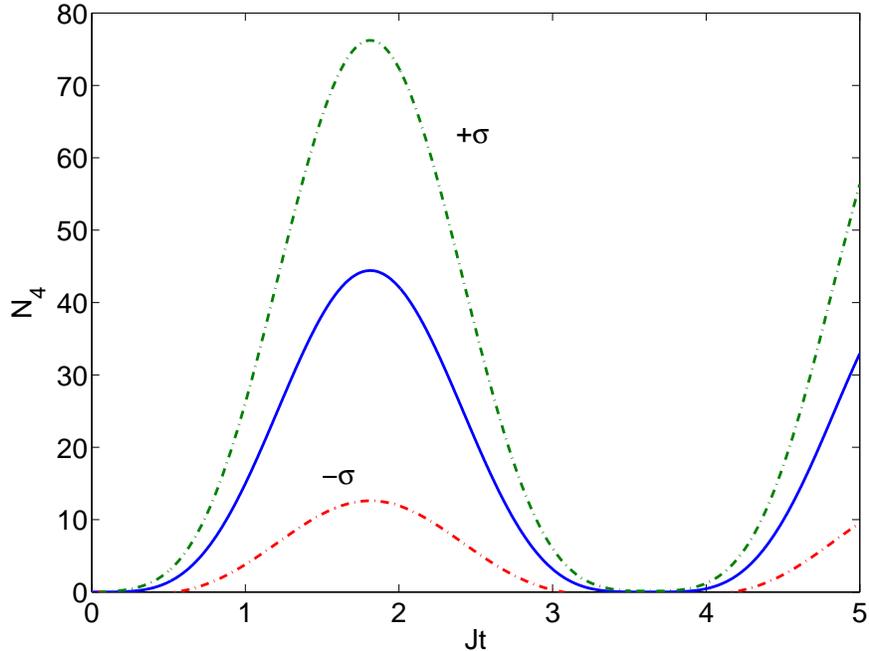}
\end{center}
\caption{(Color online) Quantum solutions for the number of atoms in the fourth well, beginning with Fock states, and the mean plus or minus one standard deviation. The initial conditions are the same as for Fig.~\ref{fig:classzero} and the solutions are the average of $3.87\times 10^{5}$ stochastic trajectories of the positive-P equations. The solutions for initial Fock states are identical for all values of $\theta_{3}$.}
\label{fig:Fzero}
\end{figure}

When we begin the simulations with Fock states of definite atom number in wells $1$ and $3$, we find that the device loses its phase dependence completely. The average results are almost identical to those found semi-classically for $\theta_{3}=\pi/2$, and do not change irrespective of the initial phase. This is not unexpected since Fock states can be roughly thought of as having indeterminate phase. As can be seen in Fig.~\ref{fig:Fzero}, however, the standard deviation around the mean solution is significant. In practice this means that the device would be totally inappropriate as a phase sensitive gate if the initially occupied wells are in number states. As this is usually thought of as the natural state for Mott insulators, this would be problematic.

Although we have considered the two extreme cases of coherent and Fock states here, it is worth noting that there are other possible initial states. As an example, the collisional nonlinearity of condensed atoms has been shown to give a ``crescent" state, where the Wigner function is stretched in the phase space~\cite{Jacob}. This initial state has been previously shown to affect the dynamics of molecular association~\cite{AMBEC2003,AMBEC2004} and would be expected to have an effect with our device as well. The advantage of the positive-P representation is that it allows us to consider arbitrary initial quantum states, as well as the calculation of any correlations that can be expressed as normally-ordered operator expectation values.

\section{Conclusions and Discussion}
\label{sec:conclude}

In conclusion, we have shown that the positive-P representation is a useful tool for analysing the performance of atomtronic devices where mesoscopic numbers of sites and atoms are involved. It is easily extended to much greater numbers of both than we have considered here. We have also shown that mean-field claculations can be very misleading for such systems and it is therefore important to undertake full quantum calculations. We have considered two markedly different initial states, showing that the coherent states, as expected, give results close to the semi-classical predictions. On the other hand, if the atoms are initially in Fock states, the semi-classical predictions are almost useless. As shown previously, correlations between the atomic modes at each site can also be calculated with this representation, which will open the door for fully quantum atomtronics in which use can be made of properties such as entanglement and EPR steering. This will definitely open up regimes that are not readily accessible with standard electronics.

\section*{Acknowledgments}
This research was supported by the Australian Research Council under the Future Fellowships Program (Grant ID: FT100100515), The New Zealand Marsden Fund and a Rutherford Discovery Fellowship from The Royal Society of New Zealand.


\begin{thebibliography}{99}

\bibitem{overview}{B T. Seaman, M. Kr\"{a}mer, D.Z. Anderson, and M. J. Holland, \pra {\bf 75}, 023615 (2007).}
%
\bibitem{MHmodel1}{M.C. Gutzwiller, \prl {\bf 10}, 159 (1963).}
%
\bibitem{MHmodel2}{J. Kanamori, Prog. Theor. Phys. {\bf 30}, 275 (1963).}
%
\bibitem{MHmodel3}{J. Hubbard, J. Proc. R. Soc. A {\bf 276}, 237 (1963).}
%
\bibitem{BHmodel}{H. Gersch and G. Knollman, \pr {\bf 129}, 959 (1963).}
%
\bibitem{Jaksch}{D. Jaksch, C. Bruder, J.I.Cirac, C.W.Gardiner, and P. Zoller,P./, \prl {\bf 81}, 3108 (1998).}
%
\bibitem{Ruschhaupt} {A. Ruschhaupt and J.G. Muga, \pra {\bf 70}, 061604 (2004).}
%
\bibitem{Micheli}{A. Micheli, A.J. Daley, D. Jaksch, and P. Zoller, \prl {\bf 93}, 140408 (2004).}
%
\bibitem{Stickney}{J. Stickney, D.Z. Anderson, and A.A. Zozulya, \pra {\bf 75}, 013608 (2007).}
%
\bibitem{Pepino}{R.A. Pepino, J. Cooper, D.Z. Anderson, and M.J. Holland, \prl {\bf 103}, 140405 (2009).}
%
\bibitem{battery}{A.A. Zozulya and D.Z. Anderson, \pra {\bf 88}, 043641 (2013).}
%
\bibitem{Gajdacz}{M. Gajdacz, T. Opatrn\'{y}, and K.K. Das, \pla {\bf 38}, 1919 (2014).}
%
\bibitem{open}{R.A. Pepino, J. Cooper, D. Meiser, D.Z. Anderson, and M.J. Holland, \pra {\bf 82}, 013640 (2010).}
%
\bibitem{DFW}{D.F. Walls and G.J. Milburn, {\em Quantum Optics} (Springer-Verlag, Berlin, 1995).}
%
\bibitem{Rab2008}{M. Rab \etal, \pra {\bf 77}, 061602 (2008).}
%
\bibitem{Bradly2012}{C.J. Bradly, M. Rab, A.D. Greentree, and A.M. Martin, \pra {\bf 85}, 053609 (2012).}
%
\bibitem{Opatrny2009}{T. Opatrn\'{y} and K.K. Das, \pra {\bf 79}, 012113 (2009).}
%
\bibitem{Pplus}{P.D. Drummond and C.W. Gardiner, \jpa {\bf 13}, 2353 (1980).}
%
\bibitem{myJPB}{M.K. Olsen, \jpb {\bf 47}, 095301 (2014).}
%
\bibitem{myJOSAB}{M.K. Olsen, arXiv:1411.4104.}
%
\bibitem{BHJoel}{G.J. Milburn, J.F. Corney, E.M. Wright and D.F. Walls, \pra {\bf 55}, 4318, (1997).}
%
\bibitem{Chianca4well}{C.V. Chianca and M.K. Olsen, \pra {\bf 83}, 043607 (2011).}
%
\bibitem{Chiancathermal}{C.V. Chianca and M.K. Olsen, \pra {\bf 84}, 043636 (2011).}
%
\bibitem{states}{M.K. Olsen and A.S. Bradley, \oc {\bf 282}, 3924 (2009).}
%
\bibitem{Steel}{M.J. Steel, \etal, \pra {\bf 58}, 4824 (1998).}
%
\bibitem{JoeSuper}{J.J. Hope M.K. Olsen, \prl {\bf 86}, 3220 (2001).}
%
\bibitem{EPRBEC}{K.V. Kheruntsyan, M.K. Olsen, and P.D. Drummond, \prl {\bf 95}, 150405 (2005). }
%
\bibitem{SMCrispin}{C.W. Gardiner, {\em Stochastic Methods: A Handbook for the Natural and Social Sciences}, (Springer-Verlag,
Berlin, 2002).}
%
\bibitem{Jacob}{J.A. Dunningham, M.J. Collett, and D.F. Walls, \pla {\bf 245}, 49 (1998).}
%
\bibitem{AMBEC2003}{M.K. Olsen and L.I.  Plimak, \pra {\bf 68}, 031603 (2003).}
%
\bibitem{AMBEC2004}{M.K. Olsen, \pra {\bf 69}, 013601 (2004).}
%


\end{thebibliography}
\end{document}